\documentclass[a4paper,onecolumn]{IEEEtran}

\usepackage{color, colortbl,url,graphicx}
\usepackage[table]{xcolor}
\usepackage{subfigure}

\newcommand{\Fig}[1]{Fig.~\ref{#1}}

\begin{document}
%
%

\title{SymbioCity: Smart Cities for Smarter Networks}

\author{Federico Chiariotti $^{1}$, Massimo Condoluci $^{2}$, Toktam Mahmoodi $^{2}$ and Andrea Zanella $^{1,*}$\\
$^{1}$ University of Padova, Padova, Italy; $^{2}$ King's College London, UK; \\
Tel.: +39-049-8277770; email: zanella@dei.unipd.it}

\maketitle

\begin{abstract}
The ``Smart City'' concept revolves around the idea of embodying cutting-edge ICT solutions in the very fabric of future cities, in order to offer new and better services to citizens while lowering the city management costs, both in monetary, social, and environmental terms. In this framework, communication technologies are perceived as subservient to the smart city services, providing the means to collect and process the data needed to make the services function. In this paper, we propose a new vision in which technology and smart city services are designed to take advantage of each other in a symbiotic manner. According to this new paradigm, which we call ``SymbioCity'', smart city services can indeed be exploited to improve the performance of the same communication systems that provide them with data. Suggestive examples of this symbiotic ecosystem are discussed in the paper. The discussion is then substantiated in a proof-of-concept case study, where we show how the traffic monitoring service provided by the London Smart City initiative can be used to predict the density of users in a certain zone and optimize the cellular service in that area.
\end{abstract}

\section*{NOTE}
This report has been submitted to the ETT Transactions on ``Emerging Telecommunications Technologies", by Wiley, and is currently under review.  
\clearpage
\setcounter{page}{1}

 \section{Introduction}
The ``Smart City''~\cite{Batty2012} paradigm extends the Internet of Things~\cite{perera2014sensing} (IoT) vision by focusing on an urban scenario in which the data collected from sensors are exploited to optimize services delivered to citizens, improve city management and quickly react to events within the city.
Today, smart city is likely the most popular buzzword among institutions, organizations, companies, and individuals concerned with the administration and the development of modern urban areas~\cite{Batty2012,Neirotti201425}.
Such a popularity roots in the huge expectations enclosed by the smart city vision, which promises to offer better services to citizens with lower costs for the public administrations and to open a multi billion market.\footnote{
Pike Research on Smart Cities: \url{http://www.pikeresearch.com/research/smart-cities.}} \footnote{
``Smart Cities Market - Global Industry Perspective, Comprehensive Analysis and Forecast, 2014--2020'': \url{http://www.syndicatemarketresearch.com/market-analysis/smart-cities-market.html} }

The Information and Communication Technology (ICT) plays a crucial role in this framework. Indeed, several ICT solutions are already exploited in many modern cities in order to make urban settings more efficient, and improve living standards. For example, pressure bands and induction coils are often placed across the main boulevards to monitor the intensity of the traffic entering and exiting the city; ``dome'' cameras are used for surveillance and traffic monitoring in main intersections or critical areas; weather stations collect environmental data in different parts of the city to monitor temperature, humidity, rain intensity, and air pollution; road signs and smartphone applications provide real-time information about available parking places and public transport services, while other apps are used for the car-sharing and bike-sharing services.

Although these applications are clearly useful to both the city managers and the citizens, they are still developed in a vertical fashion, relying on dedicated technologies and lacking interoperability both at the communication and service layer. 
This fragmentation hinders the full development of the smart city vision, which rather calls for an integrated approach where heterogeneous technologies and services become fully interoperable, following the IoT paradigm~\cite{perera2014sensing}.
The smart city scenario is indeed a main use-case for IoT technologies~\cite{zanella2014internet}, and has contributed to fuel the great excitement on this subject, which has been gathering more and more attention in the last years from both the academic and industrial communities, as testified by the explosive growth in the number of studies, products and technologies related to the IoT world~\cite{10.4108/eai.26-10-2015.150599, 10.1109/ACCESS.2016.2573678}. 

\begin{figure}[htbp]
	\centering
	\includegraphics[scale=0.4]{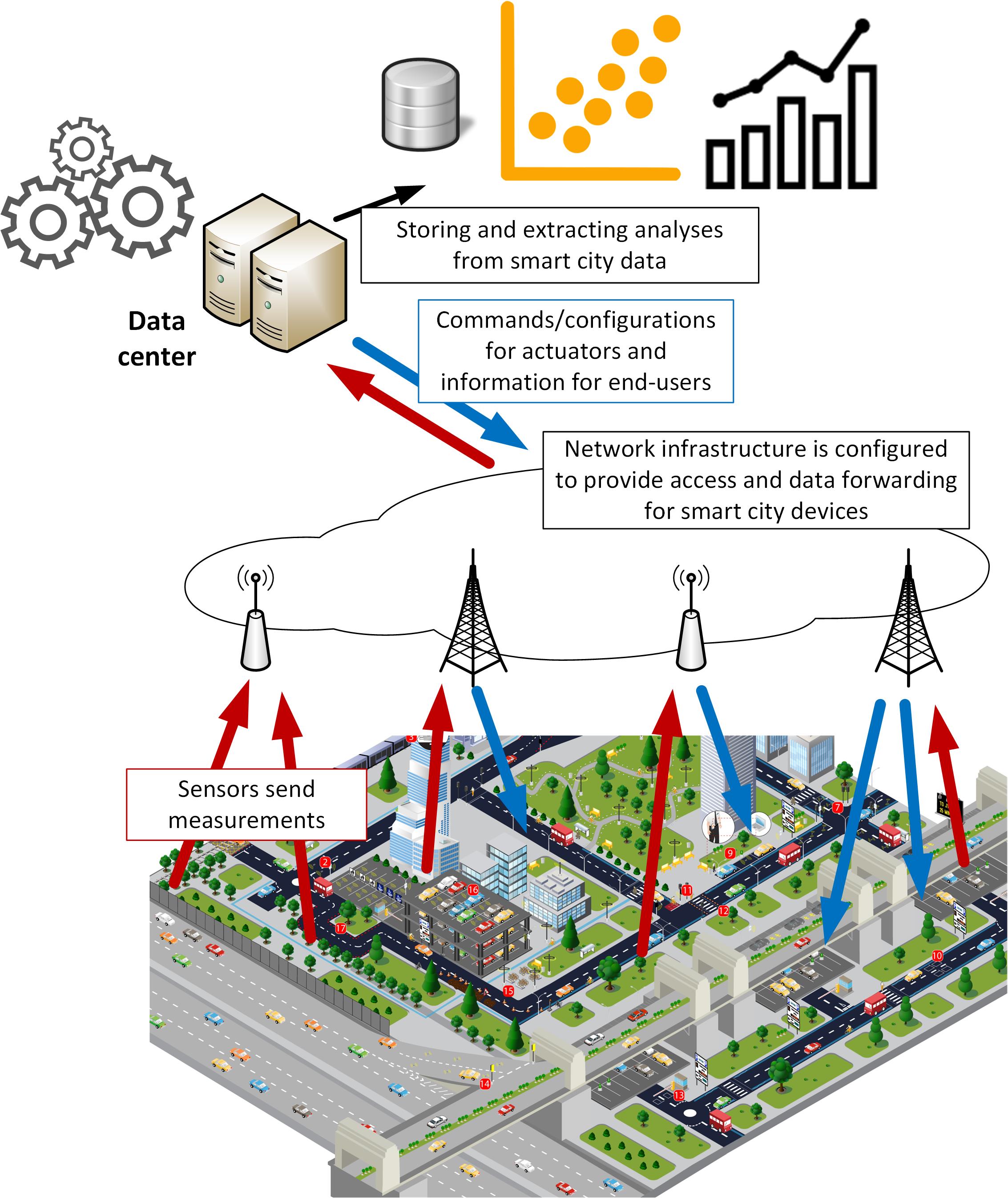}
	\caption{A representation of a smart city environment which highlights the role of ICT infrastructure in providing connectivity to smart city devices.}
	\label{fig:smart_city}
\end{figure}

Therefore, the ICT and, more specifically, the IoT technologies are key ingredients of the smart city scenario, basically representing the nervous system of the complex organism that is embodied in the smart city concept. The role of ICT is indeed to provide a capillary sensing system that covers the urban area, capable of generating and reporting information regarding the physical and social activities in the city and to implement the actions decided by the smart city services to improve the quality of life of the citizens and the efficiency, equity, and sustainability of the city management.
Nonetheless, the current vision of the Smart City paradigm conveys an inner imbalance between the role of ICT and Smart City services, in that the first should be designed and developed to sustain the second, as exemplified in \Fig{fig:smart_city}. 

In this paper, instead, we advocate that ICT and smart city services are \emph{symbiotic} and that a synergic design of the two systems can yield mutual benefits. In our vision, the roles of ICT and Smart City services are rebalanced: if on one hand the communication systems will make it possible to collect data and perform actions as dictated by the smart city services, on the other hand the information provided by such services can be used to improve the performance of the communication systems . For example, the information regarding the traffic conditions in the city during the day can be used to adapt the cellular coverage of the area, e.g., turning on/off secondary base stations according to the expected density of customers, or to pre-fetch popular content at the network edge to decrease the data traffic that crosses the core network and reduce the latency experienced by the final user. Information regarding public events, like open-air concerts, demonstrations, or political rallies, which can be made accessible through a smart city platform, can be exploited  to adapt the resource management algorithms used by the cells that serve the event venue by, e.g., increasing the uplink capacity in prediction of the massive upstreaming of multimedia content generated by the people attending the event. In a nutshell,  the richer context information that can be obtained from the Smart City services can be exploited by the network operators to improve the quality of service offered to all customers, thus increasing the efficiency of the communication systems.

To the best of our knowledge, this change of perspective is new and does not belong to the (common) interpretation of the Smart City paradigm. 
Therefore, we proposed to name this new concept as \emph{Symbiocity}, to remark the symbiotic and synergic interaction between smart city services and underlying enabling technologies. 

In the rest of this paper we will provide a quick overview of the services and the enabling technologies that are combined in the ``classical'' smart city vision, and we will then elaborate more on the possible synergies between them. We hence proposed a simple but enlightening case-study to exemplify the inner potential of the proposed approach. More specifically, we show that the information on vehicular traffic in a certain area (which we obtain from real data traces collected in the city of London) can be exploited to improve the resource allocation between human-type and machine-type traffic sources, showing significant performance gains with respect to a standard, traffic agnostic resource management algorithm, in terms of aggregate throughput enjoyed by the users, packet delivery probability, and latency for delay-critical applications. Despite the simplicity of the use-case, and the preliminary nature of the analysis and results, still the study show the capabilities of the Symbiocity concept. 

The remainder of the paper is organized as follows. Section~\ref{SCS} presents the main services of a Smart City, while Section~\ref{ET} expands on the enabling technologies. We discuss the evolution of the interaction between smart city and ICT and describe our vision for an integrated SymbioCity in Section~\ref{SV}. We present a proof-of-concept application of the SymbioCity paradigm in vehicular networks in Section~\ref{results}. Finally, Section~\ref{conclusions} concludes the paper.

\section{Smart city services}\label{SCS}

The exploitation of sensors and actuators together with a modular event-based structure (such as the one presented in~\cite{filipponi2010smart}) can be used to introduce a large variety of novel services in a city.
Such services range from human-oriented services targeting a better quality of life for people living in a smart city, to machine-oriented services handling the management of the smart city itself~\cite{zanella2014internet}. In this section, we provide a quick overview of possible smart city services that, besides contributing to the ``smartness'' of the city, can also be of use for improving the performance of the communication services in the city, thus becoming part of our SymbioCity vision. Here we describe the services from the traditional smart city perspective, leaving  the analysis of their interplay with the enabling technologies to Section \ref{SV}.
\smallskip \\\emph{Crowd mapping.} One important piece of information for city managers is the flow of people in the different parts of the city during the day. Crowd mapping can be obtained, for example, by processing the images collected by dome cameras, or from the data gathered by road-traffic sensors, or from the load of public WiFi hotspots and cellular base stations.  The analysis of people flows can reveal which areas of the city are the most or least popular at any hour, and this information can be of use to adjust city development plans, monitor social habits, improve safety in certain areas by increasing the surveillance or promoting social events, and so on.

demand-response functions aimed at improving the comfort of the people in the structure while minimizing the energy consumption. Typical examples include dimming lights based on the natural lighting and human presence in the rooms, controlling Heating, ventilation and air conditioning (HVAC) to guarantee comfort while reducing the power consumption, planning the activation of electrical appliances to smooth peaks in the power demand, and so on. 	\smallskip \\\emph{Logistics.} This service tracks items through their whole life chain and plans deliveries and waste management. It often combines different types of technologies: i) Radio Frequency IDentification (RFID) technologies, used to speed up the enumeration/identification of items in containers and in stock, ii) various environmental sensors, used to check whether the item has always been kept within specific conditions, and iii)  vehicle tracking systems, such as {Global Positioning System (GPS) or cellular-based localization techniques to track the path followed by the items. \smallskip \\\emph{Public transport/taxis.} This service provides information about the position and status of public buses and taxis to commuters in order to compute the best routes or to cut waiting times. In addition, public transport or taxi companies could exploit data coming from their vehicles to properly manage their fleets (e.g., check the status of the vehicles, re-optimize routes). This type of service requires technologies for tracking the location of the public vehicles in the city and predicting their trajectories, as well as sensors to monitor the load of the buses (i.e., the number of passengers), in order to identify hotspots for passenger loading, alighting, and exchange. 
\smallskip \\From this quick overview, two things are apparent: first, the implementation of smart city services relies on a number of different communication systems and technologies, some of which will be deployed ad hoc, while others are already in place to provide more traditional services; second, most of the smart city services gather a lot of information on the activity of humans and machines. This information can be very precious for the optimization of the very same communication networks that should provide these and other services. In the next section, we will focus our attention on such enabling technologies and their current limits.


\section{Enabling technologies}\label{ET}

With an abstraction effort, we can decompose the technology architecture needed to sustain the smart city services in the following basic building blocks:
\begin{itemize}
\item \emph{Hardware}: peripheral devices (sensors and actuators) that generate data and execute actions and commands, physically interacting with the environment.
\item \emph{Access technologies}: transmission technologies (mainly wireless) that are used to interconnect the peripheral devices with special units, often called collectors or \emph{gateways}, which provide the point of contact with the Internet. There is a plethora of such access technologies, either specifically designed for smart city purposes or designed for general purpose and customized to be utilized in the smart city. The use of hybrid access for extending capacity and coverage~\cite{10.1109/ICC.2014.6883806} is one on the main trends in wireless networks, delivering smart city services. We elaborate on the access technologies further in this section.
\item \emph{Network technologies}: protocols and platforms to manage the communication between the gateways and the rest of the world (e.g., cloud services) through the Internet. Notable representatives of this type of technologies are the MQ Telemetry Transport (MQTT) and the Representational State Transfer (ReST) protocols, often combined with suitable ontologies (see, e.g., \cite{filipponi2010smart,10.4108/eai.26-10-2015.150599}). Another example of networking technologies is the use of network virtualization in deploying multiple services over a shared infrastructure, like multi-tenancy~\cite{condoluci2015cyclon}.
\item \emph{Data analytics}: algorithmic techniques to extract useful information from the data generated by the peripheral devices ~\cite{jin2014information, Rathore201663}.
\item \emph{Decision making}: strategies to determine suitable course of actions based on the information provided by the data analytics techniques.
\end{itemize}

In the rest of this section we will focus on the access and network technologies that, in our opinion, would gain the most from the symbiotic combination with smart city services. The smart cities services will be largely based on existing communication infrastructures, which have been deployed for other uses and will have to make space for these new services and applications. As of today, the wireless access to sensor nodes scattered over the city is mostly provided by cellular systems, which were originally deployed to offer traditional human-based voice and data services. However, some new technologies, specifically designed for IoT services, may untangle human-type and machine-type communications. 
The access technologies that are currently used in the smart city context can be roughly divided into the following three main families:
\smallskip \\\emph{Cellular systems.} The almost ubiquitous service coverage and the commercial and technological maturity of cellular systems make them a natural solution to provide connectivity to IoT end-devices. Indeed, many telecom operators include in their commercial offer bundles for machine-to-machine (M2M) data traffic aiming to this new type of services.
However, current cellular network technologies have been designed for traditional human-initiated wideband services, which are significantly different in terms of traffic demands from smart city services, and as the volume of M2M traffic grows the issues caused by this are becoming evident~\cite{Biral20151}.
\smallskip \\\emph{Short-range multi-hop technologies.} Short-range technologies have almost complementary  characteristics with respect to cellular technologies: explicitly designed for short-range M2M communications, operating in unlicensed Industrial, Scientific and Medical (ISM) frequency bands, and offering a very limited coverage area (a few meters).  Examples of standards in this category are 802.15.4~\cite{15.4} from Institute of Electrical and Electronics Engineers (IEEE), Bluetooth Low Energy~\cite{BLE}, and Z-Wave~\cite{g.9959}. Most short-range technologies actually support multi-hop packet delivery, but the management of these so-called mesh networks can be tricky and appears particularly impractical when extended to a wide urban area. Nonetheless, short-range technologies are still of interest in the smart city scenario to provide local coverage (e.g., within buildings) or access to the main delivery network by means of opportunistic point-to-point relaying.
\smallskip \\\emph{Low-Power Wide-Area (LPWA).} LPWA technologies have been recently proposed as the ultimate solution to provide data access to the IoT peripheral nodes. Specifically designed for M2M connectivity, they generally provide very low bitrates, low energy consumption, and wide geographical coverage. Some relevant LPWA technologies are LoRaWAN, Sigfox, Ingenu~\cite{s16050708,centenaro2015long}. The evolution of such technologies has so far followed a path parallel to that of the mainstream cellular systems, though the next generation (5G) of the global wireless communication technologies envisions a converge of basically all services into a common platform. As of today, LPWA networks are growing in popularity and commercial interest, though the technology is still in the midst of the rush to become the de-facto standard for machine-type communications. Very recently, the arena has been enriched by the new proposal from the 3rd Generation Partnership Project (3GPP), which has finally released the specifications of the Narrowband Internet of Things (NB-IoT) technology. Although being the last arrived, NB-IoT can take advantage of the widespread presence of the existing cellular infrastructure, which offer an important competitive gain. In any case, these technologies are quite constrained in terms of transmission capacity and the perspective of massive deployment calls for the study of more advanced and context-aware management protocols to limit mutual interference and performance degradation.  \medskip

Despite the number and variety of access technologies that are currently available to support smart city services, each of them may potentially reveal limits in case of massive deployment of the envisioned services. Many of these shortcomings, however, can be avoided or mitigated by employing different kinds of network optimization techniques. In the remaining of this section we briefly discuss what we believe are the most interesting of such techniques and which are their current limitations.  
\smallskip \\\emph{Cell breathing.} Green communications and networking have become an important topic of research in the past few years; one of the most effective schemes to reduce the environmental footprint of the network is cell breathing, i.e., shutting down cells with low load. Furthermore, adapting the cell coverage is a way to control the interference and trade connection speed for cell capacity.
\smallskip \\\emph{HetNets.} A promising approach to satisfy the increasing demand for high-speed wireless access is to place pico and/or femto base stations within a macro cell, a paradigm known as Heterogeneous Networks (HetNets). However, the deployment of HetNets raises new problems. For example, as networks tend to become more dense, handovers become more frequent and need to be carefully managed to avoid load imbalances or resource starvation~\cite{guidolin2014markov}. Therefore, the HetNet technology has the potential to increase the capacity of the communication system in a certain area, but also requires more sophisticated management mechanisms to fully exploit its potential.
\smallskip \\\emph{Context-aware content distribution.} The massive diffusion of smartphones and tablets has contributed to the increase of the demand for mobile multimedia content. This demand can be further exacerbated by the diffusion of video surveillance services at the urban level. Among the different techniques proposed to address this challenge, a promising approach consists in proactively caching the most popular content in different locations, closer to the final users, thus reducing the latency and the traffic over the core network. To maximize these performance gains, however, content caching should account for the users mobility and the nature of the events that generate the traffic demand.
\smallskip \\\emph{Cloud Radio Access Network (Cloud RAN) and Coordinated Multi Point (CoMP).} Cloud RAN \cite{checko2015cloud} is a new concept in cellular network management, moving most of the packet processing to the cloud. This shift makes it easier to coordinate between base stations and avoid inter-cell interference in dense networks~\cite{irmer2011coordinated}. The advances in Cloud RAN is also an enabler for Downlink and Uplink Decoupling (DUDe) within dense and heterogenous networks~\cite{GC2016}, as expected in smart city.\smallskip \\\emph{Device to Device (D2D), Vehicle to Vehicle (V2V), Vehicle to Infrastructure (V2I) and Infrastructure to Vehicle (I2V).} The strictly hierarchical structure of traditional wireless networks can be augmented by direct communications between either IoT objects or traditional Internet devices. Again, for these approaches to be effective, it is essential to have some information about the traffic patterns~\cite{pappalardo2016caching}.\smallskip \\\emph{Software Defined Networking (SDN) and Network Function Virtualization (NFV).} These techniques can be used to control the structure of the network and adapt it to changing requirements, giving a higher priority to emergency services and redistributing the load across the backhaul links~\cite{10.1109/MVT.2014.2333765, wwrf32}. The SDN and NFV paradigms can also be used to differentiate the service offered to M2M traffic, which typically requires low delay and jitter and high packet delivery ratio, and human-type traffic, more sensitive to bandwidth and delay.\smallskip \\\emph{Access \&\ scheduling protocols.} Massive access is going to be a problem for the current cellular networks because of the predicted surge in M2M traffic \cite{Biral20151,polese2016m2m}. This problem calls for new access schemes and scheduling mechanisms that need to predict, or at least quickly react to, the waves of access requests from multiple devices. A possible approach is to dynamically adapt the settings of some protocol parameters to avoid network collapse. Other 5G applications, such as low delay transmission or the services envisioned in the so-called \emph{tactile Internet}, also need special solutions due to the very restrictive constraints~\cite{fettweis2014tactile}. These advanced access and scheduling protocols need to be enhanced with context awareness and predictive capabilities.

\emph{Multihoming.}
In most mobility scenarios, handovers and changes in the cell load can make the capacity experienced by a mobile user suddenly change; the integration of different access technologies and the appropriate protocols can provide users with a smoother experience while allowing the network to rebalance the load~\cite{gladisch2014survey}.
\smallskip \\\emph{Network Slicing.} Different services within the smart city ecosystem would also require different solutions in terms of user mobility support, handover management, multimedia management (proactive content caching), storing, routing, multihoming, broadcasting, etc. One of the solutions that can provide multiple features over a single network is network slicing~\cite{NGMN5GWP,EW2016}. Through network slicing, functionalities such as specific mobility functions or anchor point migration, will be configured according to different types of provided services. In addition, other aspects such as path configuration and load balancing need to be carefully set in order to handle smart city services while maintaining a high degree of freedom in handling mobility issues.

\section{From smart city to SymbioCity}\label{SV}

The first part of this section presents the existing trends in ICT towards a smarter and more dynamic adaptation of the smart city services to the requirements. The latter part, however, will present the SymbioCity paradigm, highlighting its novelty and how it can be integrated in today's vision of smart cities.

\subsection{ICT changes to support smart city services}

One of the major current trends in networking is to increase context awareness, and using such awareness for self-organization and self-optimization. This trend has been backed up by number of changes in the design and developments of the underlying communications platform, that are detailed in the previous section (also summarized in Table \ref{fig:smart_city_table}). For example, in Self-Organizing Networks (SON), the ICT infrastructure can react to variations of data traffic by changing its configuration and improving the connectivity of end-users accordingly (seen in Fig.~\ref{fig:son}).

In SON~\cite{7508902,7813753,7498098}, the core and access network monitor some key performance metrics (such as inter-cell interference, traffic load, packet error rate, etc.) and automatically optimize the network configuration. Alternatively, and in a more device-centric fashion, the sensors nodes can keep track of the performance of their connection (e.g., transmission failure rate) and tune some parameters~\cite{zanellaGC2016}, such as data compression level, and back-off period. Both schemes are context-aware, and they rely on the context being \textit{network performance}. Our SymbioCity vision, however, introduce context-aware optimization for the network, where the context is enriched by the data collected from the smart city services (in our use-case, city traffic planning).

\begin{figure}[htbp]
	\centering
	\includegraphics[scale=0.4]{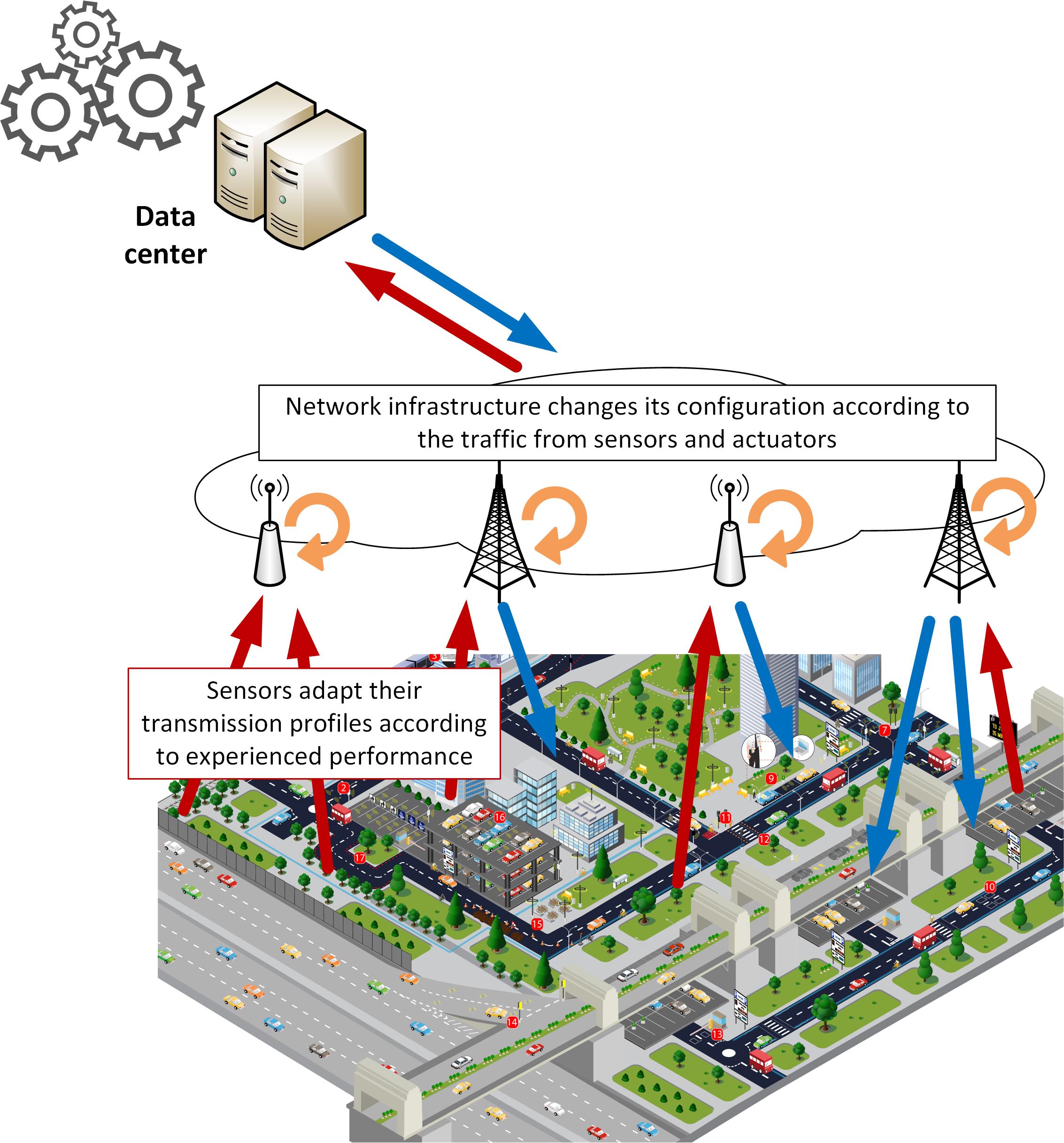}
	\caption{Evolution of the solutions in the ICT to dynamically and more effectively handle smart city services.}
	\label{fig:son}
\end{figure}

\subsection{The SymbioCity vision}
As mentioned earlier, the effectiveness of most network optimization techniques depends on the capability of the system to dynamically adapt its behavior to the variations of the operational context. Unfortunately, inferring the operational context is a complex process, in particular when observations are restricted to the only parameters internal to the communication system. On the other hands, many smart city services embody precious information regarding the operational scenario, which could be exploited to improve the performance of the communication system. In the following of this section, we further elaborate on this concept which is depicted in Fig.~\ref{fig:symbiocity}.

In our vision, the network infrastructure collects information on its performance under different configurations and under load, but also enriches it with the information provided by the smart city. For example, expected mobility and network usage patterns can be extracted from the statistics on road users or public transport users. Such enriched context is enabled through the symbiotic relation between smart city and the communication/ICT infrastructure. Exploring this new and enriched context, and its potential in improving network optimization either through SON or other Quality of Service (QoS) delivery, is the main focus of this paper.

\begin{figure}[htbp]
	\centering
	\includegraphics[scale=0.35]{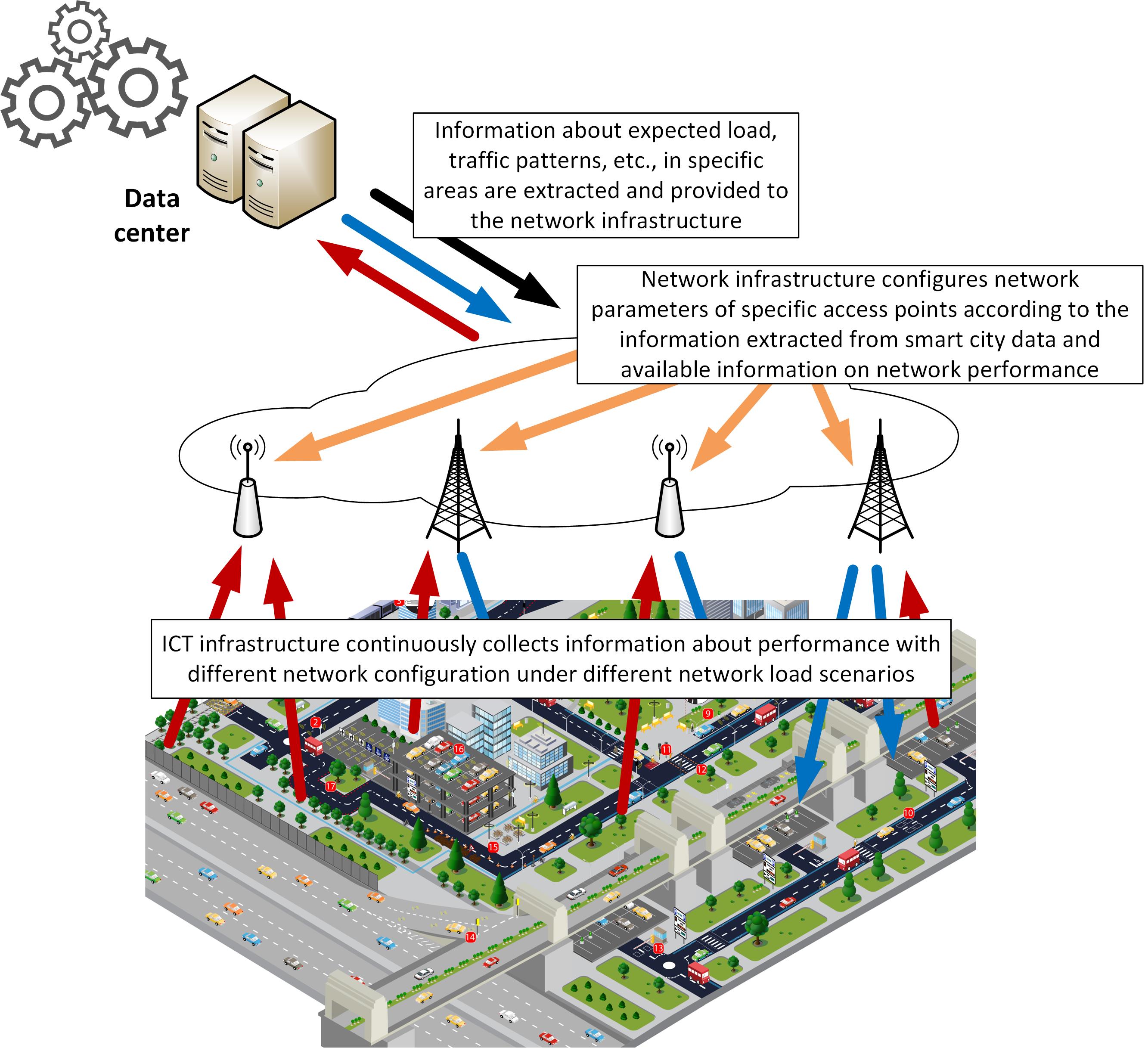}
	\caption{The SymbioCity vision where knowledge extracted from smart city data is exploited by ICT infrastructure to configure itself according to traffic patters expected for smart city services.}
	\label{fig:symbiocity}
\end{figure}



\begin{table*}[htbp]
	\centering
	\caption{Table of the possible usefulness of smart city parameters for network optimization.}
	\label{fig:smart_city_table}
	\resizebox{\textwidth}{!}{\begin{tabular}{l|lllllllllllllll}
\multicolumn{1}{c|}{\textbf{\large Smart City Services}		}	&\multicolumn{10}{c}{\bf \large Smart network techniques}
		\\
		\cline{1-12}
 \multicolumn{1}{c|}{} & \multicolumn{1}{c}{\it  SoN} &\vtop{\hbox{\strut \it Cell}\hbox{\strut \it breathing}} & \vtop{\hbox{\strut \it HetNets}\hbox{\strut \it management}} & \vtop{\hbox{\strut \it  Context-aware}\hbox{\strut  \it content distr.}} & \it  Cloud RAN (CoMP) &  \vtop{\hbox{\strut  \it D2D, V2V,}\hbox{\strut  \it V2I, I2V}} &  \vtop{\hbox{\strut \it  SDN, NFV}\hbox{\strut  \it (e.g. emergency)}} & \vtop{\hbox{ \it Access \&\ }\hbox{\strut  \it scheduling}} &  \it Multihoming &  \vtop{\hbox{\strut  \it Service}\hbox{\strut  \it resilience}} & \vtop{\hbox{\strut  \it Network}\hbox{\strut  \it Slicing}}\\
\cline{2-12}
		Crowd mapping				& & strong & & & strong & strong & & strong & strong & strong & \\
		\vtop{\hbox{\strut Traffic map and}\hbox{\strut traffic light control}} & & & strong & medium & strong & strong & & medium & medium & medium & medium\\
		Parking monitoring			 & medium & weak & & & & strong & & & medium & medium & \\
		Building automation			&medium & weak & & medium& & & &medium & & & strong\\
		Logistics				& strong &  & strong & strong & & & & & & & \\
		Public transport/taxis				& medium & & medium & medium & & medium & weak & & strong & strong \\
		\vtop{\hbox{\strut Patrolling}\hbox{\strut (dome cameras)}}				& & medium & medium & & & & & & & & strong & \\
	Bike/car sharing				& medium &  & & & & medium & weak & & medium & medium & \\
		City event calendar			& strong & strong & & strong & & & medium & & & strong & medium\\
	\end{tabular}}
\end{table*}


Table~\ref{fig:smart_city_table} shows a summary of the possible synergies in a fully developed SymbioCity.  The networking applications of the smart city data range from SoN to service resilience in stressful situations such as big events and natural disasters.

The first and foremost type of smart city data is \textit{location}: smart city services can help the network gauge the position of both human-type and machine-type users, allowing it to turn on and off secondary base stations (cell breathing), coordinate interference among micro- and femto-cells (Cloud RAN and CoMP), better control the handover process of the mobile users, and support both massive access for M2M traffic and low-delay applications (Access \& scheduling), i.e., augmented reality overlays and the tactile Internet~\cite{fettweis2014tactile}.

Knowledge of vehicular traffic patterns and the traffic light system can even enable a predictive approach, in which the network preventively counteracts situations at high risk of congestion. For example, pico base stations can be placed in the areas that are usually interested by traffic jams (e.g., upon the traffic lights of main roads) and activated in case of road congestion, in order to absorb the increase of data traffic due to the large number of vehicles in the area and the tendency of the passengers to kill the time surfing the web or playing online games while waiting for the green light.

On a longer timescale, the city's event calendar can make the network aware of large concentrations of people in advance, allowing service providers to take measures to mitigate the interruption of service that has plagued such events since mobile phones first became popular. For example, information regarding a public event that will collect a large number of people on a certain area of the city as, e.g., a concert in a square or stadium, can trigger SON mechanisms to cope with the massive uplink traffic generated by the crowd (for example, by increasing the bandwidth reserved for the uplink channel in the cells that cover the interested region), or can trigger a reconfiguration of the forwarding rules in the transport network to enable flawless streaming of the event to remote users.

Another interesting development is the integration of location, traffic, parking and public transport data to enable a true city-wide vehicular network. The high speed and unpredictability of cars have always made long-distance vehicular communication extremely complex, but recent efforts to integrate Vehicle to Vehicle (V2V) and Vehicle to Infrastructure (V2I) communication have had some success, and a smart city-based approach to the problem might be the last piece of the puzzle. Parked cars, buses and strategically placed access points, connected to the Internet with fiber optic links can form the backbone of such a vehicular network, and smart city data are necessary to keep it from collapsing (D2D, V2V, I2V). City infrastructure, public transport, taxis and car sharing services can also be exploited to give users multiple paths to the Internet, giving them more resilience to channel outages and a more robust transmission~\cite{gladisch2014survey}.

Some of these factors are already taken into account by network planners, and some dynamic ad hoc adjustments are already widely used in cellular network management. However, the lack of a systematic infrastructure makes these optimizations haphazard and extremely dependent on human intervention. In our vision, the SymbioCity should be able to integrate far more diverse data, in real time and without human intervention. While the actual optimization logic will be completely automated, its goals will not: in an environment as complex as a city, network and city planners can aim at very different objectives, even varying them dynamically. For example, in a very polluted city, it might be wise to optimize the network to be more eco-friendly and reduce power usage, while resources might be better used by increasing vehicular communication during rush hour.

A modular, fully interoperable structure should also allow adding new data and applications with ease: since the city grows and changes to adapt to its population, the SymbioCity should grow with it, adapting its functions to the changing environment.



\begin{figure*}[htbp]
	\centering
	\includegraphics[scale=0.3]{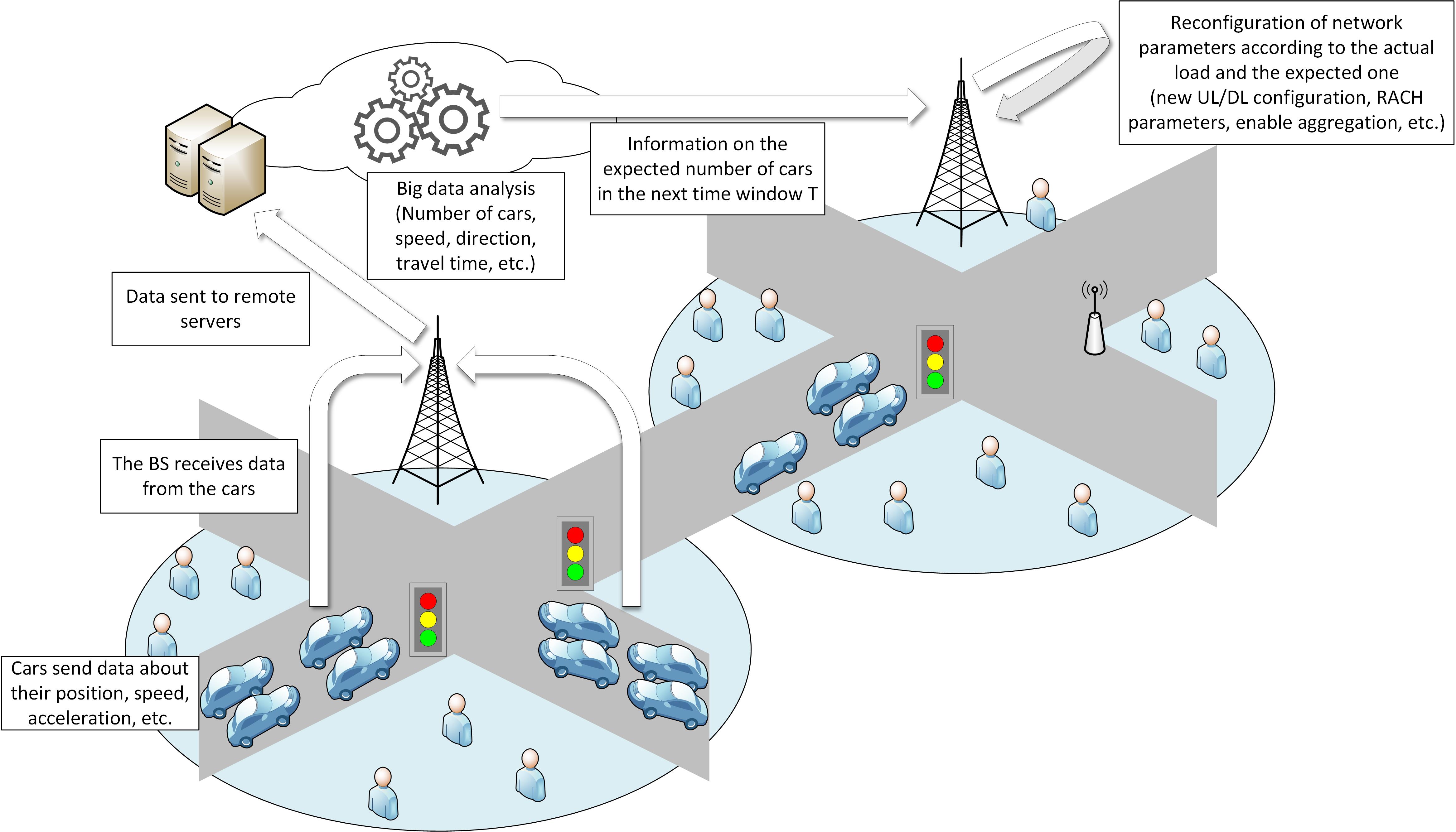}
	\caption{SimbioCity in a vehicular application: adaptation of network parameters of a base station serving a junction according to the data collected from the cars.}
	\label{fig:smart_city_adaptation}
\end{figure*}
\section{Performance evaluation}\label{results}

In this section, we test the SymbioCity concept in a practical scenario. We remark that the purpose of this analysis is to showcase the potential benefit of using Smart City data for network optimization in a simple but enlightening case-study, rather than developing a new solution to cell congestion in a smart city.  Targeting this objective, we consider vehicular communications within a smart city environment and present some results that show how the information on vehicular traffic intensity can be exploited to activate ``smart network'' optimization techniques with the aim of maintaining a certain level of QoS for human-type (HT) and machine-type (MT) communications despite the variable density of users in the area.
Fig.~\ref{fig:smart_city_adaptation} illustrates a possible way SymbioCity could improve network performance in such a scenario: the data received from the cars for drive-assistance services is also used by the SymbioCity engine to estimate the amount and typology of data traffic that can interest a certain geographical area. The configuration of the base stations or other network elements in the area can then be adapted to better support the communication.

\begin{figure}[!ht]
	\centering
	\includegraphics[scale=0.3]{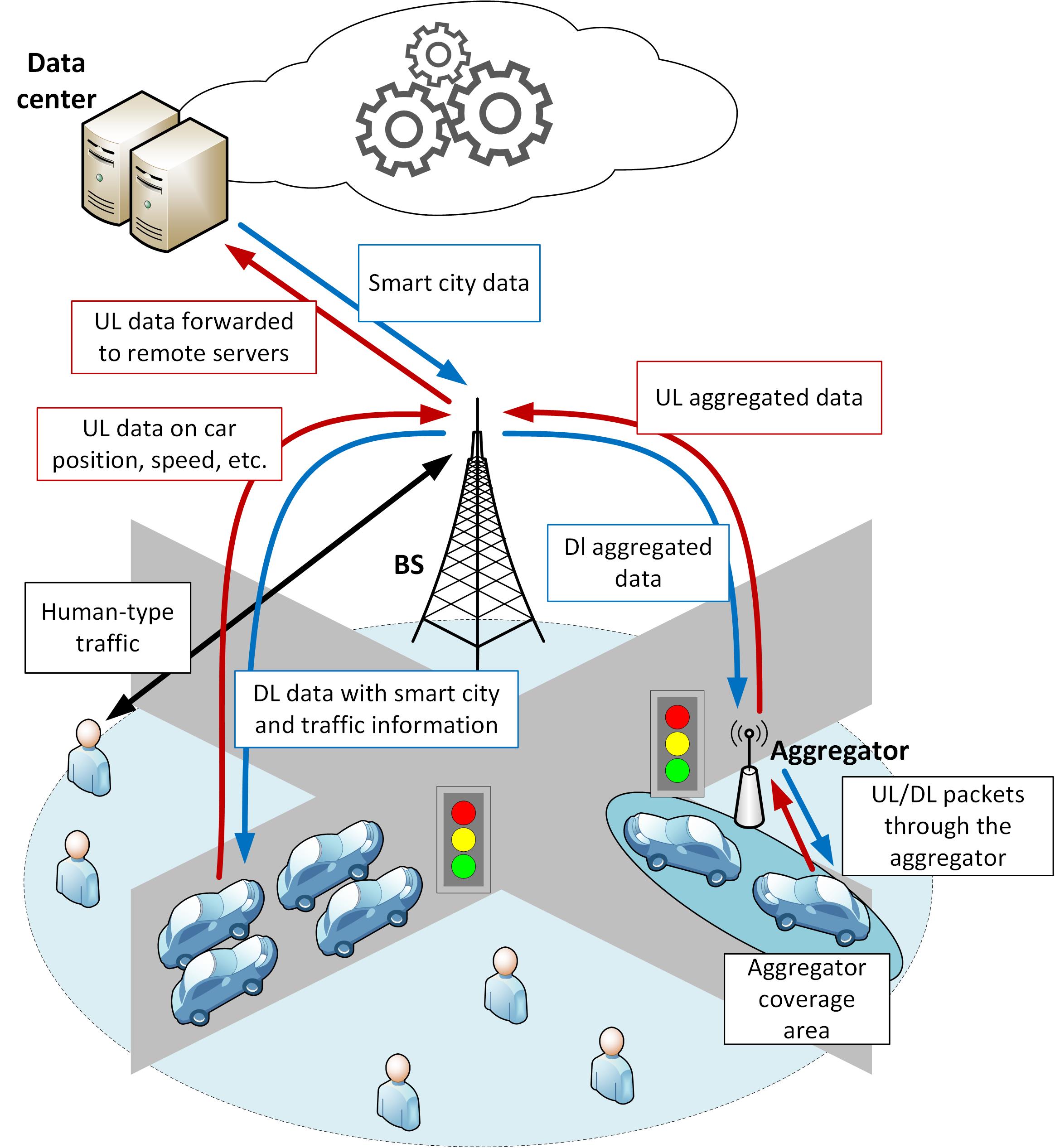}
	\caption{Testing scenario: one junction with cellular communications handling human-type and vehicular traffic.}
	\label{fig:IM_smart_city}
\end{figure}

\begin{figure}[!ht]
	\centering
	\includegraphics[scale=0.35]{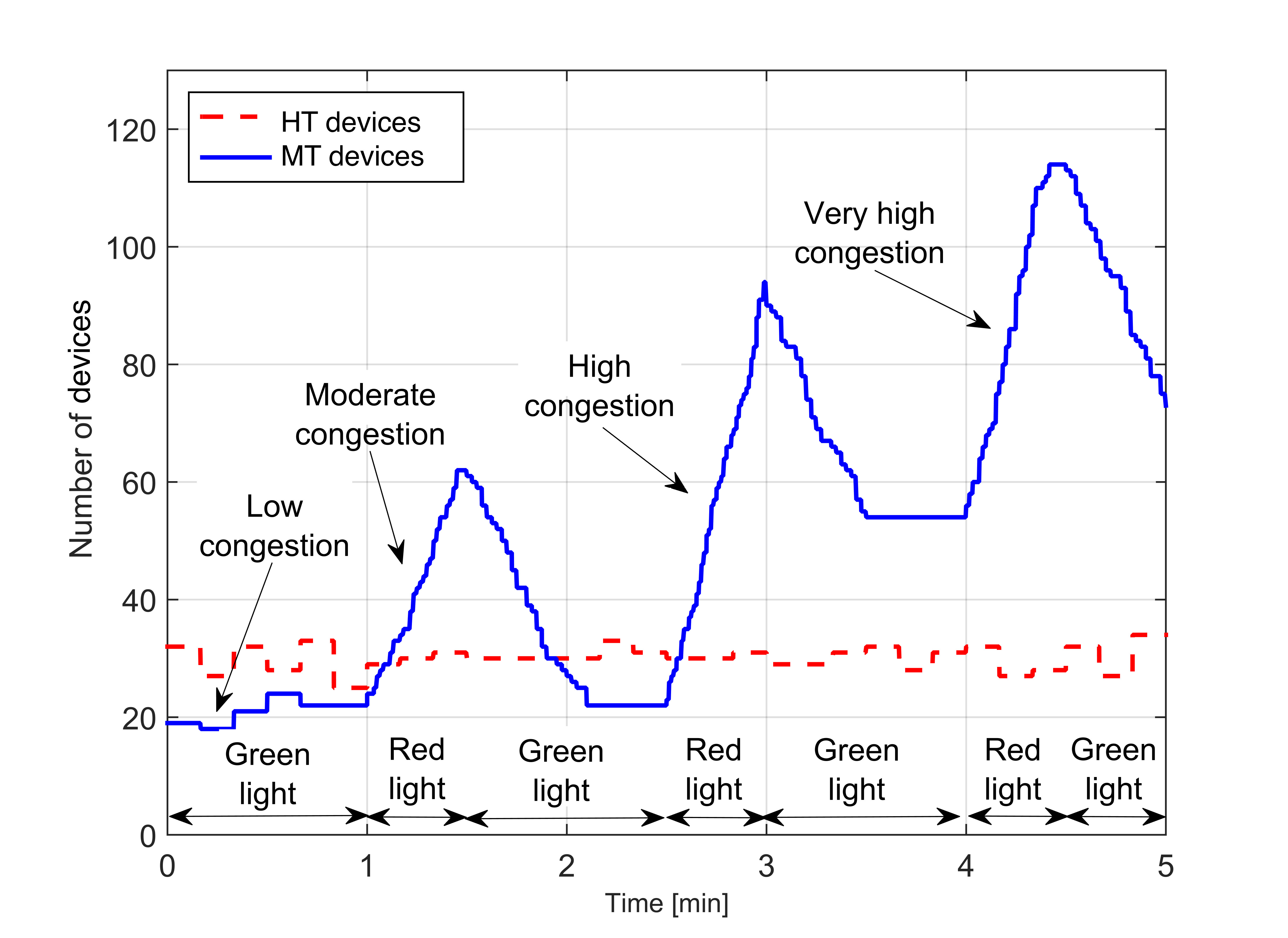}
	\caption{Traffic to be managed by a base station by considering human-type and car devices in a junction.}
	\label{fig:traffic}
\end{figure}

We analyze the simple but representative use case depicted in \Fig{fig:IM_smart_city}, where one junction has a central Long Term Evolution (LTE) evolved NodeB (eNB) serving both vehicular and human-type mobile devices. Using  the real data traces provided by Transport for London (TfL),\footnote{https://tfl.gov.uk/} a prototype smart city platform deployed in the city of London, we analyzed the road traffic conditions of several junctions and defined four reference levels of road congestion based on the mean number of vehicles that, by considering a traffic light period, are on the crossing roads within a distance of about 500~m from the center of the intersection, i.e., on an area roughly equal to the coverage range of the LTE base station located in proximity of the center of the junction.

The London traffic sensor network is composed of thousands of sensors, detecting presence in the intervals of 250 ms. By checking the time evolution of these sensors, and identifying the gaps between vehicles, we can estimate both the number of cars crossing an intersection in a given period and their average speed. This estimation is extremely simple, and can easily be automated.

The road congestion levels are computed based on the data from TfL. From the information collected by the monitoring system, we found a mean duration for the green light signal of 1 minute, and a red light duration of 30~s; which represent a reasonable timing for a busy junction. Then, denoting by $n$ the number of static or mobile vehicles in the intersection during a green-red interval (90 seconds), we identified four cases:\footnote{Note that this congestion thresholds are arbitrary, though reasonable for the considered scenario.}
\begin{itemize}
	\item \emph{Low congestion}: $n\leq 20$;
	\item \emph{Moderate congestion}: $20<n\leq 60$;
	\item \emph{High congestion}: $60 < n \leq 100$;
	\item \emph{Very high congestion}: $n>100$.
\end{itemize}
We hence consider a time window of 5 minutes, with the traffic pattern depicted in Fig.~\ref{fig:traffic}. At any point in time, half of the cars are moving at a constant speed along one of the two roads (i.e., the road with green light), while the other half are waiting for the light to turn green.

We assume each car exchanges UDP packets of 800 bytes with a drive-assistance server located somewhere on the Internet. The uplink packets are sent every 100~ms, while downlink transmissions are spaced apart by 1~s; the traffic asymmetry models a service that needs to collect information about cars' status, like speed,  direction, diagnostic measurements, and so on, rather frequently, while it sends non-critical information to the vehicle (e.g., road congestion, traffic, parking availability, etc) less often. Besides MT traffic generated by the drive-assistance service, we also assume a certain volume of human-type (HT) traffic. We distinguish two scenarios with regard to the HT traffic: in the first, the passengers in the cars do not generate any data traffic, and the HT traffic is generated by 30 static users who are randomly placed in the buildings surrounding the intersection. In the second scenario, 25\% of the cars have a passenger that generates data traffic, in addition to the 30 static users of the first scenario.
The HT generated by each of these devices is modeled as an asymmetric flow, with an average bitrate of 800~kbit/s in downlink and of 400~kbit/s in uplink.

The simulation was set up in the network simulator \emph{ns-3}~\cite{riley2010ns}, using the LENA+ module that provides a more realistic Random Access Channel~\cite{polese2016m2m} for LTE connections. Additional simulation parameters are listed in Table~\ref{table:param} \footnote{In the 802.11n, the $2.4$~GHz ISM band is chosen for maximum coverage, despite being more congested. Also, legacy 20 MHz channel is considered sufficient for the M2M traffic to be aggregated and relayed via LTE.}

We consider as a baseline a standard network configuration where the base station has a channel bandwidth of 5\,MHz, corresponding to 25 LTE resource blocks (RBs), which is a configuration commonly adopted by mobile operators. This reference scenario will be referred to as \emph{Standard}. Note that, with the Standard setting, the capacity of the LTE cell would be sufficient to accommodate the HT traffic. However, the presence of even a relatively small number of MT devices may result in some performance loss, as will be discussed later.

We hence consider two possible network optimization strategies that can be adopted by the SymbioCity engine to improve the QoS of MT and HT data flows when the road congestion level and the data traffic offered to the LTE cell increase.

The first technique consists in exploiting V2V communications to collect the traffic generated by the cars waiting for the red light  on a concentrator node that, in turn, forwards the aggregated data to the service center using a V2I (LTE) link. The distributed configuration requires no infrastructure, and several distributed protocols have already been proposed in the literature on vehicular networks~\cite{raya2006, fathollahnejad2013reliability, hagenauer2014advanced, abrougui2012efficient}. The aggregation alleviates the massive access problem that may be experienced by the LTE cell when a large number of vehicles is at the junction, each trying to establish a connection with the LTE base station to send their MT packets. The massive number of requests coming from the vehicles may impact the HT flows, resulting in a degradation of their QoS. This technique shall then be triggered when congestion is moderate to high, since the overhead of electing a concentrator and relaying the aggregate MT traffic to the LTE base station would not be justified in case of light traffic. This optimization technique will be indicated in the following as \emph{Aggregator}.

The second optimization strategy consists in increasing the capacity of the cell that serves the junction by doubling the number of available RBs. In a practical deployment, this bandwidth increase may be achieved, e.g., by powering a twin base station that is activated only if needed, to save operational costs, or alternatively by ``stealing'' underutilized bandwidth from the adjacent cells. This optimization technique, which is indicated in the following as \emph{Extra Resources}, clearly has a dramatic impact in terms of performance gain, but its cost makes it convenient only when the other techniques are not longer effective, i.e., in case of very high congestion.

\begin{table}[!ht]
	\centering
	
	\caption{Simulation parameters} 
	\label{table:param}
	
	\begin{tabular}{lc}
	\hline
	\hline
		Parameter 					&	Value 			\\
		\hline
		Downlink carrier frequency 			& 945 MHz			\\
		Uplink carrier frequency			& 900 MHz			\\
		RB bandwidth					& 180 kHz			\\
		Available bandwidth				& 25-50 RBs 			\\
		Number of eNBs					& 1				\\
		eNBs beamwidth					& $360^{\circ}$ (isotropic)	\\
		TX power used by eNBs 				& 43 dBm 			\\
		eNB noise figure				& 3 dB 				\\
		LTE scheduler					& PF/FF				\\
		Number of HT devices			& 30-35-45-55-60				\\
		Number of MT devices			& 20-60-100-120			\\
		HT requested traffic (downlink)		& 800 kb/s			\\
		HT requested traffic (uplink)		& 400 kb/s			\\
		HT packet size				& 1000 B			\\
		MT requested traffic (downlink) 	& 6.4 kb/s	\cite{6515060}		\\
		MT requested traffic (uplink)		& 64 kb/s	\cite{6515060}		\\
		MT packet size			& 800 B	\cite{6515060}			\\
		Vehicle speed					& 5 m/s				\\
		Vehicle inter-arrival time			& 50 ms				\\
		Aggregator standard				& 802.11n			\\
		Aggregator carrier frequency			& 2.4 GHz 			\\
		Aggregator bandwidth				& 20 MHz	\\
		Aggregator maximum TX power			& 30 dBm			\\
	\hline
		
	\end{tabular}
	
\end{table}

\begin{figure*} [htbp]
	\subfigure[First scenario\label{fig:thr_HTC_time_first}]
	{\includegraphics[scale=0.095]{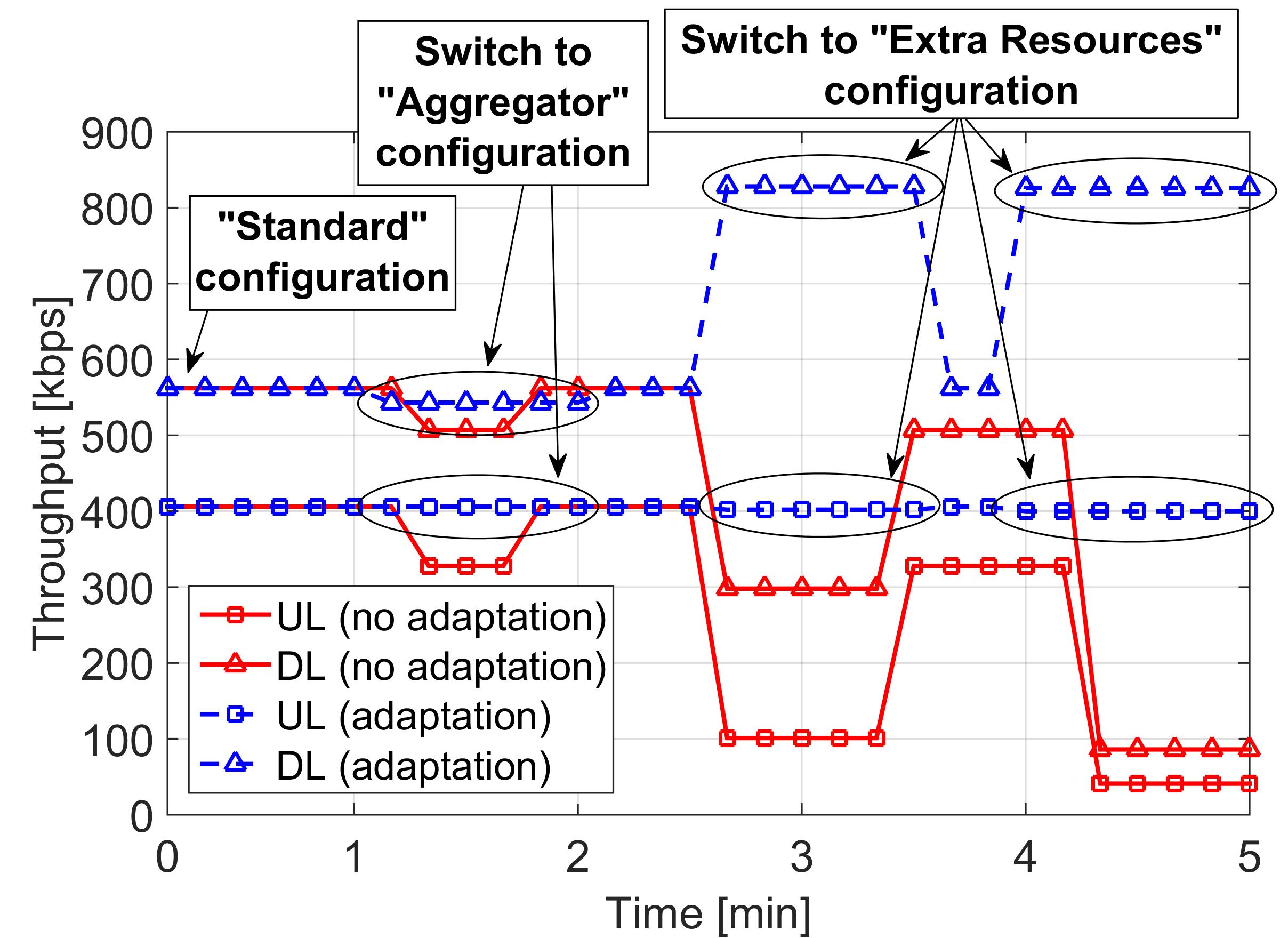}}
	\hspace{0.2cm}
	\subfigure[Second scenario\label{fig:thr_HTC_time_second}]
	{\includegraphics[scale=0.095]{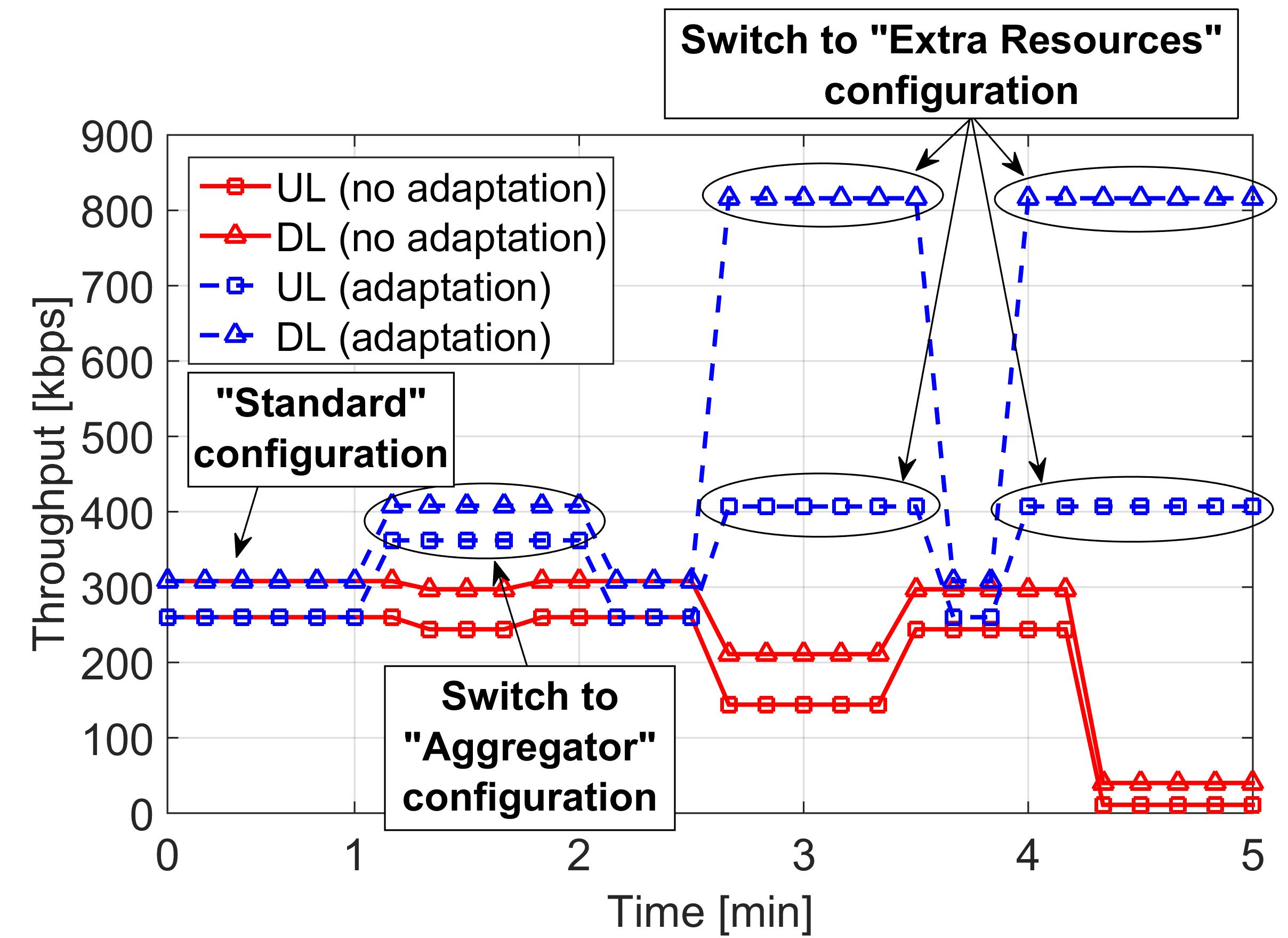}}
	\caption{Throughput variation experienced by HT users with and without adaptation of system parameters.}
	\label{fig:thr_HTC_time}
\end{figure*}

\begin{figure*} [htbp]
	\subfigure[First scenario\label{fig:del_MTC_first}]
	{\includegraphics[scale=0.1]{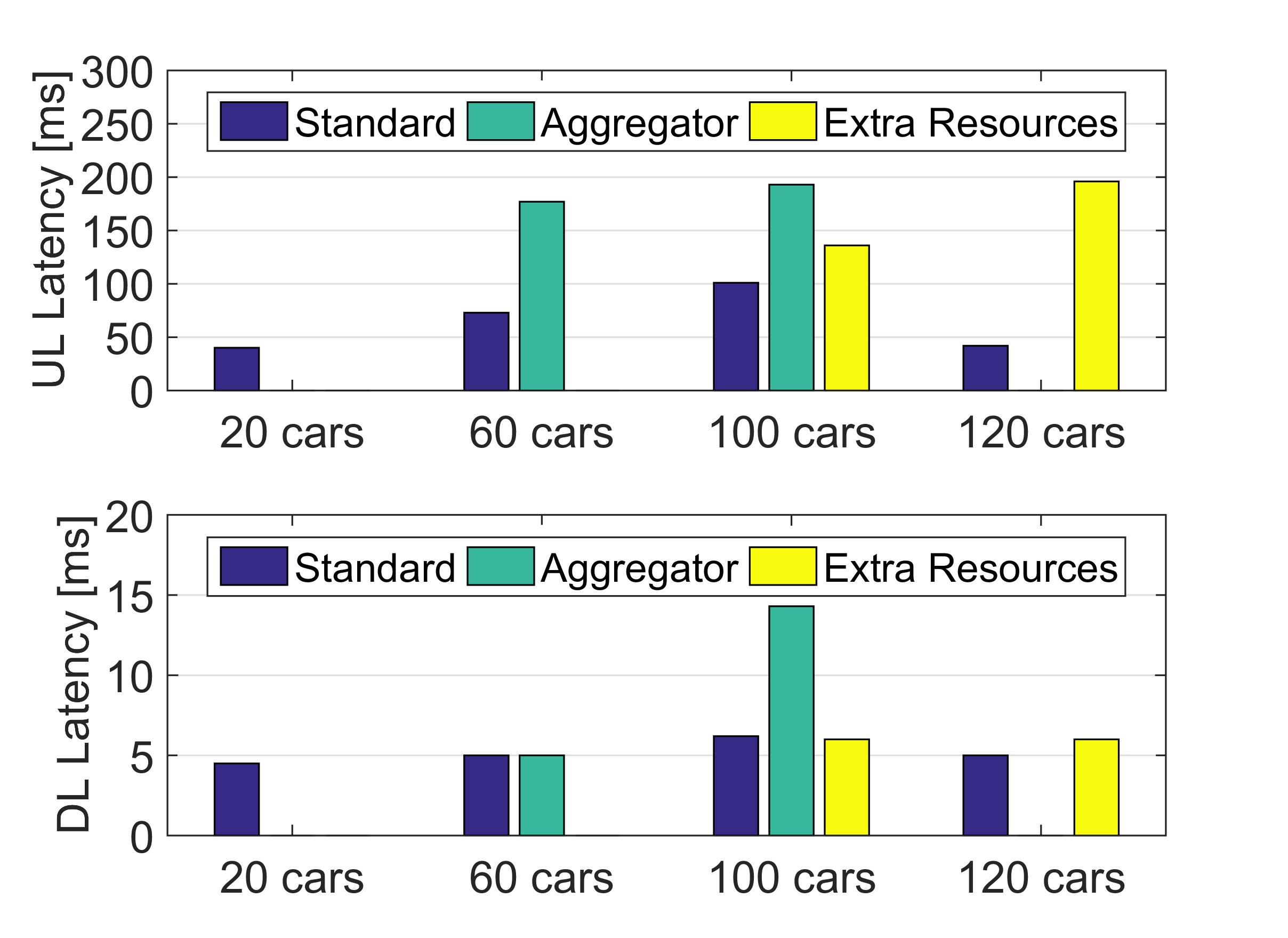}}
	\hspace{0.2cm}
	\subfigure[Second scenario\label{fig:del_MTC_second}]
	{\includegraphics[scale=0.1]{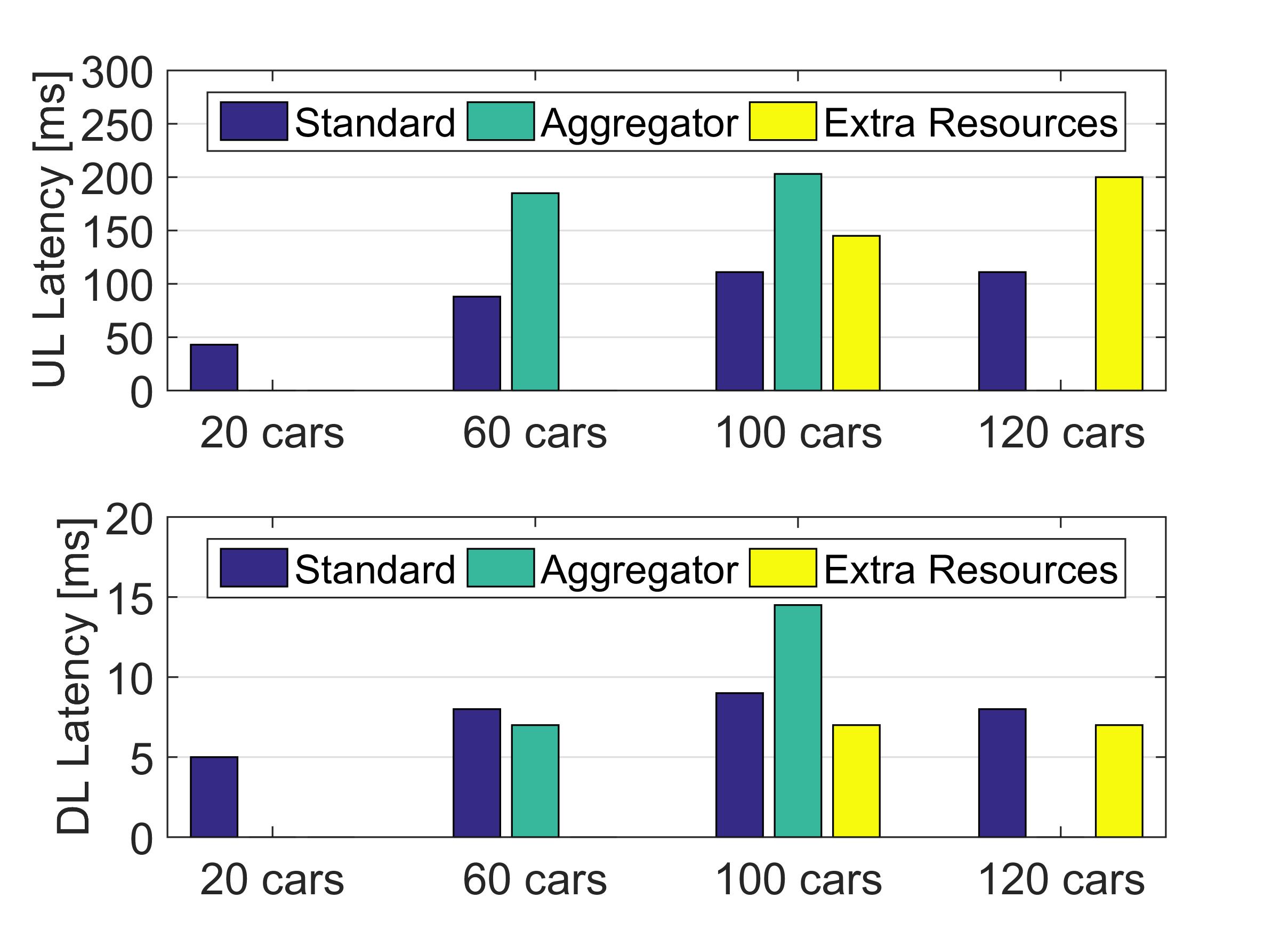}}
	\caption{Average latency experienced by MT traffic when varying the system setting and the road congestion level.}
	\label{fig:del_MTC}
\end{figure*}

Fig.~\ref{fig:thr_HTC_time} shows the uplink (UL, triangle markers) and downlink (DL, square markers) throughput of HT users in the time window shown in \Fig{fig:traffic} when considering the {Standard} setting for the communication network (red solid lines), and when applying the {Aggregator} technique for medium road congestion levels and the {Extra Resources} intervention for high and very high congestion levels. We can observe that, under the Standard system configuration, the presence of even moderate MT traffic is enough to lower the downlink throughput of HT devices by about 30\%, while the (lower) uplink throughput is not affected by a low MT traffic. A deeper analysis of the traffic patterns shows that the throughput drop of HT traffic is mainly due to the Proportional Fair Scheduling policy adopted by the base station~\cite{7508441} which, in case of scarcity of transmission resources, tends to equally divide the channel capacity among all the backlogged flows, so that the largest flows get more penalized.

When the road congestion increases, both uplink and downlink throughput of HT connections sharply decreases with the Standard setting. In this case, the {Aggregator} technique can partially isolate the HT traffic from the interference of MT packets, making it possible to maintain basically the same performance experienced with Standard setting in the light congestion case. In this case, the Proportional Fair LTE scheduler does not penalize the HT traffic because a good part of the MT traffic will be carried through the single LTE connection of the aggregator node, thus fairly competing with HT traffic for channel access.

When the road congestion is high or very high, the throughput of HT flows drops significantly. The {Aggregator} technique can only partially reduce the performance loss by aggregating the MT traffic of the vehicles that are queued at red light, but the vehicles that cross the junction will still maintain a direct connection with the base station, thus interfering with the HT traffic. In this situation, the only possibility to maintain (or even improve) the QoS of HT traffic without excessively throttling the MT flows is to increase the capacity of the cell, e.g., by adopting the Extra Resources approach. Doubling the number of resource blocks that can be utilized by the base station makes the cell capacity sufficient to serve all the traffic generated in the area, so that the downlink and uplink throughput of the HT nodes reach their nominal values of 800 kbit/s and 400 kbit/s, respectively.

The price to pay for isolating the performance of HT services from the impact of MT traffic is, of course, a degradation of the performance of this last category. Fig.~\ref{fig:del_MTC} is a bar plot with the mean delay experienced by MT packets in uplink and downlink directions, when increasing the road congestion levels. The results for the {Aggregator} and the Extra Resources techniques have been reported only for the congestion levels where such techniques can be reasonably adopted.

Since the MT downlink traffic is very low, the latency is also low and does not vary significantly with the road congestion level. Instead, it is interesting to observer that the {Aggregator} scheme increases the delays of MT packets both in uplink and downlink, due to packet relaying and to the Proportional Fair scheduler, which provides fair resource sharing among all backlogged nodes, irrespective of the number of queued packets. This factor needs to be considered even in the low-congestion scenarios, although it might be overcome by using a different scheduling policy.
Another effect of the scheduler is that the packets from the moving cars are prioritized over the aggregated ones, which makes sense in a vehicular network, since moving cars need up-to-date information more than the stopped cars~\cite{chen2010broadcasting}.

The latency of the uplink MT traffic with the Standard setting increases with the road congestion level, to decrease again in the very high congestion case where, however, a large fraction of the packets are not delivered because of the severe channel congestion, so that the few packets that the devices manage to send experience a low delay. Finally, the uplink latency when using the Extra Resources technique is relatively large in the high and very high congestion scenarios, but most of the packets are delivered to the drive-assistance service.

We note that the HT throughput and MT latency show the same trends in both the considered scenarios, i.e., with and without HT traffic generated by vehicles. The additional HT traffic has clearly a negative impact on the performance, but the drawbacks and benefits of each scheme remain the same. Moreover, we observe that, in thus study, the triggering of the Aggregator scheme has been designed to maximize the coverage and limit the interference on surrounding APs. However, further study is need to adapt the triggering to the different situations, in order to optimize the communication. The same consideration holds for the Extra Resources scheme, which has non-negligible costs in terms of management and spectrum leasing. Therefore, there is still a large space for improvements. 

This simple analysis illustrates some important facts that justify the proposed SymbioCity approach: first, the performance of the communication network and, consequently, the quality of the services delivered to the users, may exhibit significant fluctuations due to the changes in the operational conditions (e.g., road traffic, public events, whether conditions, and so on); second, the smart city services can be exploited to recognize and, to some extent, predict these changes; third, smart network optimization techniques can be employed to compensate the performance degradation in critical situations; and, finally, the performance gain is achieved only when all these mechanisms are well orchestrated and combined in a symbiotic manner.
%
\section{Conclusions} \label{conclusions}
While the Smart City vision is mainly developing the concept of using ICT solutions for delivering smart and efficient services, the new technology-enabled furniture are also great source of information for improving the ICT services. In this paper, we propose a new vision where technology and Smart City services are designed to take advantage of
each other, in a symbiotic manner. According to this new paradigm, which we call ``SymbioCity'', the wealth of sensing and measurements available through smart city are exploited to provide better connectivity services, and to optimize the overall delivery of communication services. 

Since one of the major components of the smart cities is the intelligent transportation system, we showcase our SymbioCity vision through analyzing the vehicle traffic data and utilizing such analytics for improving the performance for the vehicular communications within the 4G/LTE network. While considering both human- and vehicle-generated data traffics, we show how smart city data can be used for optimal configuration and design of the communication protocols and architecture. In order to have realistic scenarios, the smart city data and vehicular traffic models are based on the smart sensors and monitoring data from the city of London. 

We believe that, although preliminary, these results show the potential of the SymbioCity concept, thus opening the way to a new approach for the design of Smart City services and ICT systems.

\vspace{6pt}

%
%


\bibliography{bibliography}


\end{document}